\documentstyle[aaspp4]{article}

\def\'#1{{\ifx#1i{\accent"13\i}\else{\accent"13#1}\fi}}

\def\ala#1{$^{#1}$}
\def\alamenos#1{$^{-#1}$}

\def\BB{\bf B}
\def\BP{Ballesteros-Paredes}
\def\diezala#1{10$^{#1}$}

\def\doceCO{\ala{12}CO}
\def\gammeff{{$\gamma_{\rm eff}$}}
\def\kms{km~s$^{-1}$}
\def\por{$\times$}
\def\prom#1{\langle #1\rangle}
\def\rhot{\rho_{\rm th}}

\def\treceCO{\ala{13}CO}
\def\umean{{\bf{u}}_{\rm{mass}}}
\def\uu{{\bf u}}
\def\VS{V\'azquez-Semadeni}

\begin{document}

\title{Turbulent Flow-Driven Molecular Cloud Formation: 
A Solution to the Post-T Tauri Problem?}

\author{Javier Ballesteros-Paredes$^{1,2}$, Lee Hartmann$^2$, and
Enrique V\'azquez-Semadeni$^1$ }

\affil{$^1$Instituto de Astronom\'\i a, Universidad Nacional 
Aut\'onoma de M\'exico \\ Apdo. Postal 70-264, 04510 M\'exico D.F.,
M\'{e}xico.  {e-mail: {\tt javier@astroscu.unam.mx}} }

\affil{$^2$Harvard-Smithsonian Center for Astrophysics, 60 Garden St.,
MS-42, Cambridge MA 02138}

\begin{abstract}

We suggest that molecular clouds can be formed on short time scales by
compressions from large scale streams in the interstellar medium
(ISM).  In particular, we argue that the Taurus-Auriga complex, with
filaments of 10-20 pc $\times$ 2-5 pc, most have been formed by H~I
flows in $\lesssim 3$~Myr, explaining the absence of post-T Tauri stars
in the region with ages $\gtrsim 3$~Myr. Observations in the 21~cm line
of the H~I ``halos'' around the Taurus molecular gas show many features
(broad asymmetric profiles, velocity shifts of H~I relative to
$^{12}$CO) predicted by our MHD numerical simulations, in which
large-scale H~I streams collide to produce dense filamentary
structures.  This rapid evolution is possible because the H~I flows
producing and disrupting the cloud have much higher velocities (5-10
\kms) than present in the molecular gas resulting from the colliding
flows.  The simulations suggest that such flows can occur from the
global ISM turbulence without requiring a single triggering event such
as a SN explosion.


\end{abstract}

\section{Introduction}\label{intro}

It has been recognized for many years that the nearest star-forming
regions exhibit little evidence for stars of ages $\gtrsim 5$~Myr, even
though one would expect that such ``post-T Tauri stars'' (PTTSs) should
be more numerous than the $\sim 1$~Myr-old  T Tauri stars.  More than a
decade ago, Herbig, Vrba, \& Rydgren (1986) stated that ``it is a
source of some unease that this large population of PTTSs has not yet
been identified.''  Since that time, a variety of techniques have been
used to search for PTTSs, including proper motion surveys, fainter
objective prism plates, CCD photometric selection.  As outlined in \S
2, none of these techniques has yielded evidence for PTTSs in any
significant numbers.

The lack of PTTSs, combined with the presence of newly-formed stars in
all substantial nearby molecular clouds, is most simply explained if
molecular clouds like Taurus come together, form stars, and disperse in
a few Myr.  However, this simple picture causes difficulties for
current theories of star formation.  Since Taurus has a spatial extent
of 20 pc, but the molecular gas has a velocity dispersion of only about
2~\kms, it is difficult to understand how such widely separated regions
produced stars almost simultaneously, i.e., how star formation is
``triggered'' throughout the cloud on a scale shorter than the crossing
(or dynamical) timescale.  In principle, a single powerful event like a
supernova explosion might trigger star formation in a molecular cloud
over short timescales, but there is no obvious candidate for this
triggering source in the case of Taurus (Elmegreen 1993a).

In addition, Taurus is supposed to be the archetype for the
``standard'' picture of isolated, low-mass star formation, in which
magnetically subcritical cloud cores collapse to form stars only after
ambipolar diffusion has removed excess magnetic flux on timescales of
order 5-10~Myr (Shu, Adams, \&\ Lizano 1987; \cite{Mouschovias91}, and
references therein).  If protostellar cloud cores are highly
subcritical, the long ambipolar diffusion timescale makes it difficult
to understand star formation events lasting only a few Myr.  A long
diffusion time also is difficult to reconcile with statistics of cloud
cores (\cite{LeeMyers99}).

Palla \&\ Galli (1997) argued that the ambipolar diffusion timescale
simply introduces an age offset. In this picture, the molecular cloud
cores which produced the present-day Taurus stars started contracting,
say, 11-12~Myr ago, so that the cores have only become supercritical
and started collapsing over the last 1-2~Myr. However, there is no
unique ambipolar diffusion timescale; it depends upon the ionization
fraction, which in turn depends upon precise conditions of shielding
(\cite{MyersKhersonsky95}), and on how subcritical the cloud core is
initially.  Thus it is difficult to understand how ambipolar diffusion
does not introduce a spread of at least several Myr into the onset of
star formation.  Moreover, as pointed out by Fiedler \&\ Mouschovias
(1993), Nakano (1998), and Hartmann (1998), strongly subcritical clouds
must be confined by external pressure to prevent disruption by
expansion.  The possibility of producing a steady external pressure in
the generally turbulent environment of molecular clouds has been
questioned recently by \BP, \VS, \&\ Scalo (1998 = \cite{BVS99}), who
suggest that the turbulent motions are not steady, and can disrupt as
well as compress clouds.  Given these problems and the evidence of the
stellar population, it appears likely that ambipolar diffusion is not a
major constraint on the timescale of star formation in Taurus, as
argued by Nakano (1998) on general grounds, and as proposed many years
ago by Shu, Adams, \& Lizano (1987) for much denser regions.

Even if the ambipolar diffusion timescale is not relevant, there still
remains the problem of triggering star formation in Taurus.  This is
possible in a dynamical scheme, where large-scale turbulent streams
collide, collecting the material they are advecting and creating
density fluctuations (\cite{hunter79}; \cite{hunt_fleck82};
\cite{hunter_etal86}; \cite{Tohlineetal88}; \cite{Elm93b}; \VS, Passot
\&\ Pouquet 1995 = \cite{VSPP95};  \cite{BVS99}; see also \VS\ et
al.  1999). In this picture the formation of molecular clouds cannot be
considered separately from the formation of their parent diffuse H~I
clouds. If molecular gas clouds like Taurus are ``assembled'' by the
convergence of higher-velocity neutral hydrogen flows, one might
explain the nearly-simultaneous star formation over larger distance
scales without invoking a single triggering event, like a SN
explosion.

In this paper we consider the dynamical, kinematic and spatial
relationship between the neutral hydrogen and molecular gas in the
Taurus region. Our analysis is based on H~I and \doceCO\ observations,
showing that in many places there is H~I which appears to be
dynamically correlated with the molecular gas, but with substantial
velocity offsets and larger velocity dispersions. The dynamical
timescales for the clouds are then given by the scale of the clouds
divided by the velocity dispersion of the external H~I, contrary to
the assumption that the internal velocity dispersion is the relevant
quantity.  We further show that these results are qualitatively
consistent with MHD simulations of the ISM, suggesting that the
molecular clouds like Taurus can be rapidly assembled, eliminating the
need to find PTTSs in the present molecular gas complex.

In \S \ref{TTauriproblem} we review the observations leading to the
post T-Tauri problem.  In \S \ref{observations}\ we explore the
dynamical relationship between H~I and CO gas in the Taurus region,
while in \S \ref{simulations}\ we show simulations of the ISM dynamics
which we use to interpret the observations in \S
\ref{turbcompresions}.  Finally, we point out limitations of our
simulations in \S \ref{limitations}, and summarize our conclusions in
\S \ref{conclusions}.

\section{Historical Context: The post-T Tauri
problem}\label{TTauriproblem}

Analyses of molecular cloud lifetimes have rarely included constraints
from their stellar populations; yet the ages of the stars produced by
these clouds provide uniquely detailed constraints on cloud ages that
are not obtainable in any other way (e.g., Hartmann et al. 1991;
Feigelson 1996).  Because of the importance of the stellar population
ages, it is worth reviewing the situation in some detail, especially
given some of the conflicting literature on the subject.  We focus on
the Taurus-Auriga molecular cloud complex, where the most detailed
observational efforts have been made.

To illustrate the problem, Figure 1~a shows the HR diagram for stars in
the Taurus-Auriga molecular cloud region, with stellar luminosities and
effective temperatures taken from Kenyon \& Hartmann (1995).  The great
majority of the stars have ages near 1~Myr; approximately half are
younger than this, while only a few stars have ages greater than
3~Myr (see Fig~1~b).

The group of stars near the 100~Myr isochrone (i.e., near the zero-age
main sequence) were discovered through early {\it Einstein} X-ray
surveys (Walter et al. 1988, and references therein).  These stars were
originally suggested by Walter et al.  (1988) as members of the missing
PTTSs population.  However, there is a striking gap in the HR diagram;
there are very few stars filling in the age range between $\sim 5$ and
$50$~Myr.  This gap makes it very unlikely that a single, reasonably
continuous star formation event is responsible for both the near-ZAMS
stars and the young T Tauri stars.  Furthermore, with such large ages
it is difficult to connect these stars with the present Taurus
molecular cloud.  Walter et al. (1988) found ages for these stars
$\sim$ 30~Myr at a modest velocity dispersion of 2~\kms, typical of
Taurus molecular gas. These stars can have traveled 60 pc from their
birth-sites during their lifetimes, 3 times the diameter of the Taurus
molecular complex.

The apparent coordination of star formation on relatively short
timescales in Taurus (and other regions) has led to several efforts to
identify hypothetical older stars which might be present but missed in
previous objective prism surveys dependent upon strong H$\alpha$
emission. Herbig, Vrba, \&\ Rydgren (1986) obtained objective prism
data covering the Ca~II H and K emission lines, which are present in
much older stars.  Indeed, Herbig et al. (1986) identified several
outlying members of the Hyades in their survey (see also Hartmann,
Soderblom, \&\ Stauffer 1987), but failed to identify any older PTTSs.
Hartmann et al. (1991) and Gomez et al. (1992) used proper-motion
surveys to try to select members of Taurus without regard to emission
line properties; again, no substantially older stars were found.

It has been suggested that recent X-ray surveys (e.g.,
\cite{Walteretal88}; \cite{Neuhauseretal95}; \cite{Wichmannetal96})
have discovered the missing PTTSs. However, the ages of the typical G
and early K stars in these samples are uncertain, because neither X-ray
activity nor Li depletion clearly discriminate between 1~Myr-old T
Tauri stars and 100~Myr-old stars.  It seems likely that most of these
objects are 30-100~Myr-old stars, given their numbers and properties
(\cite{Bricenoetal97}).  More recent measurements of Li depletion
(\cite{MartinMagazzu98}) and limits on the M star component of the G-K
ROSAT survey population (\cite{Bricenoetal98}) support the idea that
most of the dispersed X-ray population is much older than 10~Myr.
Indeed, it is difficult to explain the spatial distributions of these
stars if they are much younger. The dispersed X-ray sources are spread
fairly uniformly over large distances (up to 70~pc in projected
distance from the center of Taurus); it would be very difficult to move
stars from the present molecular regions this far at typical velocity
dispersions of $\sim$ 2~\kms\ in anything less than about 30~Myr.
Instead, it seems much more likely that these stars are mostly $\sim
50$~Myr-old objects which formed in a variety of individual regions
whose molecular gas has by now dispersed (\cite{Feigelson96}). Whatever
the exact age of the Einstein- and ROSAT-discovered young stars, they
do not seem to represent the ``missing'' stars in Taurus of ages
3-10~Myr.

Another approach has been to consider whether the ages of the Taurus
stars have been systematically underestimated. For example, the
evolutionary tracks of Swenson et al. (1994) give ages for Taurus stars
that would be several times larger than those of D'Antona
\&\ Mazzitelli (1994), which were used in Figure~\ref{figuralee}.
However, as pointed out by Stauffer et al. (1995), if one modifies the
calibrations of evolutionary tracks so that the calculations match the
zero-age main sequence, most of the age discrepancy between those works 
disappears. Another
possibility was raised by Hartmann \&\ Kenyon (1990), who suggested
that the effects of accretion might produce spurious age estimates, but
more recent investigations (Hartmann, Cassen, \&\ Kenyon 1997; Siess,
Forestini, \&\ Bertout 1997) indicate that T Tauri accretion has little
effect, making the stars appear slightly older, not younger. One may
also note (\cite{Mizunoetal95}) that most of the pre-main sequence
stars in Taurus lie within 1-2 pc of dense molecular gas; at a velocity
of 1-2~\kms, this suggests that these stars have not been dispersing
for more than 1-2~Myr from their natal material, in agreement with the
conventional HR diagram ages.

Thus, all the Taurus observational constraints are consistent with a
picture in which the molecular gas comes together and forms stars in $\lesssim
3$~Myr, so that few stars
of ages 3-10 Myr are expected. The widely-distributed X-ray sources
mostly represent stars which, though relatively young, are generally
substantially older than 10~Myr, and it is likely that their natal
molecular clouds no longer exist.

A general absence of PTTSs implies not only that clouds form stars
rapidly, but that star-formation timescales are also relatively short.
Since Taurus is still forming stars, its ultimate dispersal time is not
known.  However, studies of other regions such as Cha I and IC 348
(Lawson et al. 1996; Herbig 1998) provide relatively little evidence
for large populations of PTTSs, especially when older ROSAT sources are
eliminated and mass-dependent biases are eliminated.\footnote{Age
estimates for cluster stars generally depend systematically on stellar
mass, probably due to birth-line errors (i.e., uncertainty in the
initial positions of protostars in the HR diagram; Hillenbrand 1997;
Hartmann 1999.)} From the absence of known molecular cloud regions
containing 10~Myr-old stars, it appears that the molecular gas may also
disperse in a few Myr, a timescale consistent with the cluster survey
results of Leisawitz, Bash, \& Thaddeus (1989).  In this picture many
molecular clouds will not have substantial populations of PTTSs.

\section{Atomic and molecular gas: Maps and velocity-position
diagrams}\label{observations}

The Taurus clouds are well-suited for understanding gas dynamics as
well as the stellar population.  The Taurus Molecular Cloud (TMC) is
one of the most well-studied molecular clouds in the sky, with
extensive H~I and CO maps; it is one of the nearest clouds and lies
well below the galactic plane, making it easier to isolate the H~I
associated with the CO cloud from the general material in the galactic
plane.  Finally, the morphology and kinematics of TMC suggests a close
relationship between the molecular and atomic gas (see PhD thesis and
series of papers by \cite{Andersson93}, and references therein).

In the following discussion we make use of both H~I and \doceCO data.
The 21 cm line data were taken from the Atlas of Galactic Neutral
Hydrogen (\cite{dap97}) obtained with the 25 m Leiden/Dwingeloo
telescope, while the \doceCO\ molecular data have been taken from
Ungerechts \&\ Thaddeus (1987), obtained with the Columbia
millimeter-wave telescope. Details of the observations and analysis may
be found in the original papers.  Here we note that both data sets have
0.5 degree spatial resolution, which facilitates comparison.  The
velocity resolution of the H~I data is 1.03~\kms, slightly lower than
that of the CO data, 0.65~\kms.  To overlap the \ala{12}CO and atomic
maps, we transformed the molecular data set (originally in {\tt
(RA,DEC)}) to galactic coordinates $(l,b)$ by triangulating and
re-sampling the transformed data in a half degree resolution grid.  We
present both sets of data extending from 140 to 201 degrees in galactic
longitude, and from $-44$ to $+17$ in galactic latitude (note that the
original \doceCO\ data does not cover the whole range. We only add
zeros in the external region, in order to match the H~I and
\doceCO\ data at large scales).

Figure~\ref{HICO} shows a velocity-integrated map in
\doceCO\ (isocontours) and H~I (gray-scale) for the velocity range
where the CO data has been obtained ($-$20.33~\kms\ $\leq v_{\rm LSR}
\leq $ 29.1~\kms). The straight lines indicate the loci of the cuts
along which the velocity-position diagrams shown in
Figure~\ref{cutposvel} are made. Assuming an optically thin medium, the
H~I intensity (in K \kms) can be considered as proportional to the
column density with a conversion factor of 1.8~\por
\diezala{18}~cm\alamenos 2.  The gray scale range in Figure~\ref{HICO}
corresponds to a column density range from
$\sim$~2.5~\por~\diezala{20}~cm\alamenos 2 (darker pixels), to a few
times \diezala{21}~cm\alamenos 2 (white pixels). A column density of
\diezala{21} cm\alamenos 2, corresponding to $A_V \sim 0.5$, is
generally taken to be roughly the minimum column density needed for
sufficient shielding of the UV radiation field to produce CO (e.g.,
Elmegreen 1993a).

In Figure~\ref{cutposvel} we present a set of six velocity-position
diagrams for both the H~I and \doceCO\ emission. The first two panels
(cuts 1 and 2 in Fig.~\ref{HICO}) are the velocity-position diagrams
for the region that formally is known as TMC, in which \treceCO\ has
been detected (see e.g.  \cite{KleinerDickman84}; \cite{Mizunoetal95};
etc.). Figure~\ref{cutposvel}~c  (cut 3) corresponds to the region
associated with Perseus Arm, and shows a strong velocity gradient, from
$\sim$ 8-10~\kms\ to $\sim - 2$~\kms.  Figure~\ref{cutposvel}~d (cut 4)
corresponds to the northernmost region mapped by Ungerechts
\&\ Thaddeus (1987). In Figure~\ref{cutposvel}~e (cut 5) we present a
velocity-position diagram that covers a large region, going from high
southern latitudes to close to the galactic plane.  Finally, in
Figure~\ref{cutposvel}~f (cut 6) we show the velocity-position diagram
for the nearby cloud L1457. In particular, note from
Figure~\ref{cutposvel}d and e that at low galactic latitudes the H~I
exhibits a large velocity spread due to the detection of material at a
wide range of distances in the galactic plane.  However, once one moves
out of the plane a sufficient distance ($b < -10^{\circ}$), the H~I
velocity width becomes substantially smaller. The H~I velocity peak is
generally fairly close to, although not identical with, the \ala{12}CO
peak. This reflects the fact that the \doceCO\ and H~I emission are
spatially and kinematically related, suggesting that they are produced
as part of a single dynamical complex (see, e.g.,
\cite{BlitzThaddeus80}).

>From Figure~\ref{cutposvel} the following features can be noted.
First, wherever there is a \doceCO\ feature, H~I is also found, with an
approximate column density similar to that required by shielding (see
above).  Second, the converse is not true: not all H~I in this velocity
system is associated with molecular gas.  Third, at a given spatial
position, the H~I emission often does not peak at the same velocity
than the \doceCO; frequently there is an offset of a few \kms\ between
the two species. Fourth, velocity widths in the H~I spectra are larger
than the \doceCO\ spectra by a factor of roughly 3 or more.  Fifth,
both the \doceCO\ and H~I line profiles are asymmetric, as indicated by
the variation of gray-scale and contours in Figure~\ref{cutposvel}a-l.
These characteristics have been noticed previously in other regions
(see e.g., \cite{BlitzThaddeus80}; \cite{MBM85};
\cite{ElmegreenElmegreen87}), and we shall focus on these properties
when comparing with numerical simulations in the next section.

\section{NUMERICAL SIMULATIONS}\label{simulations}

\subsection{The Model}\label{model}

To interpret the significance of the observed features of the atomic
and molecular gas enumerated at the end of \S {\ref{observations}}, we
turn to numerical simulations of the interstellar medium by Passot,
\VS, \&\ Pouquet (1995 = \cite{PVSP95}). These two-dimensional
simulations represent the behavior of a one square kiloparsec of the
ISM centered at the solar galactocentric distance. The simulations
solve the MHD equations, including self-gravity, parameterized cooling
and diffuse heating, the Coriolis force, large-scale shear, and
parameterized localized stellar energy input due to ionization
heating.  The parameterized cooling is as in \cite{chi_breg88}, who
fitted piecewise power laws to the standard cooling calculations of
\cite{dalg_Mc72} and \cite{ray_etal76} (see \VS, Passot, \&\ Pouquet
1996 = \cite{VSPP96} for details).  As discussed in \cite{VSPP96}, the
cooling and diffuse heating time scales are in general much shorter
than the dynamical time scales, implying that the flow is always in
thermal equilibrium, except in the vicinity of star formation sites. We
refer to the reader to \cite{VSPP96} for further details in the
equations and the model itself, and to the video accompanying
\cite{PVSP95}, which gives an animated view of the dynamics of the ISM
and shows the transient character of the clouds\footnote{See also {\tt
http://www.astroscu.unam.mx/turbulence/movies.html}}.

Because the present MHD simulations do not include chemistry, we
adopt a picture in which molecular clouds are the ``tips of the
icebergs'' of the density features in the simulations. We define a
`molecular' cloud as a connected set of pixels with density above some
density threshold $\rhot$. The selection of this density threshold
is arbitrary, but it must be a compromise between realistic values
for molecular clouds and the limitations of the simulations. Because
the densest and smallest structures in the simulations reach values
$\rho \sim$ 50~-~100~cm\alamenos 3, we select a density threshold of
35 cm\alamenos 3, unless otherwise stated. For comparison, typical mean
densities in molecular clouds start at roughly 20~cm\alamenos 3 (e.g.,
\cite{Blitz87}). In particular, an average value of 15 cm\alamenos 3 is
reported for Rosette Molecular cloud (\cite{Williamsetal95}).

The simulations have a mean density of $\prom n = 1$~cm\alamenos 3,
which corresponds to a mean column density of $\prom N \sim 3.1 \times $
\diezala{21}~cm\alamenos 3 integrating through the 1~kpc box. This value
is large enough to envision enough shielding to UV radiation to form
molecular hydrogen (see e.g. Franco \&\ Cox 1986).

For the analysis here, we use a run presented in Ballesteros-Paredes, 
V\'azquez-Semadeni \&\ Rodr\'iguez (1997 = \cite{VBR97}) called run
28.800, which has a resolution of  800 $\times$ 800 pixels. In order to
mimic as much as possible the conditions in Taurus, we turn off the star
formation at $t = 68.9$~Myr (5.3 code units). This allows us to increase
the density contrasts, and to avoid the heating from massive stars,
which are not present in Taurus. Then, we follow the time evolution of
a relatively small cloud, which is localized far away from the big cloud
complexes formed in the simulations.

\subsection{Results}\label{results}

In Figure~\ref{nube_evol} we show a time-sequence of a subsection of
the whole box of the simulation, starting 7.02~Myr after the time in
which we turn off the star formation. The grey scale indicates the
density field, ranging from 0.31 to 56.6~cm\alamenos 3, and arrows
denote the velocity field in the reference frame of the cloud. The
snapshots are separated by $\Delta t = $ 0.39~Myr (0.03 time code
units). The black isocontour is the region where the density values are
higher than 35~cm\alamenos 3. Note that the cloud is located where the
velocity field converges, as required by the continuity equation. Also,
one observes rapid growth (within less than 2 Myr) of a ``cloud'' above
our density threshold which is approximately 20 pc in length (each
pixel unit is 1.25~pc), comparable to the length of the main Taurus
clouds. 

To show that the density features observed in the simulations are
produced by the confluence of external large-scale streams\footnote{By
large-scale streams we refer to random motions whose scales are
comparable with the scales of the observed structures.} rather than
other way around (e.g., the velocities being the  consequence of, say,
gravitational collapse of the cloud), we calculate the time evolution
of both characteristic lengths and energies for the density features in
Fig.~\ref{nube_evol}.  We select a connected set of pixels with
densities equal or higher than the half value of the maximum density at
each timestep to calculate properties for the ``cloud'', a choice
motivated by the fact that the maximum density in the box is changing
substantially during the simulation (from $\sim 16$ to $\sim 60$
cm\alamenos 3).

In Fig.~\ref{longitudes} we display the evolution of two different
characteristic lengths: a) $l_x$, the maximum $x$-length of the cloud
(solid line); and b) $l_J$, the Jeans length (dotted line) calculated
as
\begin{equation}
l_J = \biggl [ {\gamma_{\rm eff} \pi c_i^2\over G \rho_0^{2-\gamma_{\rm
eff}}} \biggr ]^{1/2},
\label{jeans}
\end{equation}
where \gammeff\ is the effective polytropic index in the simulations
resulting from equilibrium between the heating and cooling rates, $c_i$
is the isothermal speed of sound, and $\rho_0$ is the mean density of
the cloud (see \cite{VSPP96} for details). The values of these lengths
are denoted on the left-hand side $y$-axis.  We also display the
evolution of three different energies: a) the absolute value of the
gravitational energy $E_{\rm grav} = -1/2 \int \rho \phi dV$
(short-dashed line), where $\rho$ is the density and $\phi$ is the
gravitational potential; b) the kinetic energy in the frame of the
cloud $E_{\rm kin,cloud} = 1/2 \int_V \rho (\uu - \umean)^2 dV$
(long-dashed line), where $\uu$ is the velocity field and $\umean$ is
the mass-weighted average velocity of the cloud; and c) the magnetic
energy $E_{\rm mag} = 1/8\pi \int_V B^2 dV$  (dot-dashed line), where
$\BB$ is the magnetic field. Note that in all the previous integrals,
the volume element is $dv = dx dy$, because of the two-dimensional
character of the simulations and thus all the energies considered here
are strictly per unit length in the $z$-direction. The values of the
energies are denoted in the left-hand side of the $y$-axis. The
$x$-axis runs from $t=0$, the timestep when we turned off the
star-formation in the simulations, to 15.2~Myrs after.

At $t=0$ the Jeans length (dotted line) is initially larger than the
$x$-length $l_x$ (solid line) (the difference was larger at earlier
times).  This comparison suggests that, at least in this direction, the
density structure is not formed by gravitational collapse. We have
calculated also the Jeans-length for the whole subregion (not showed
here), and find that it is again larger than the length of the
subregion size.  These results suggest that even if the region must
`feel' the action of the gravity, it is Jeans stable at the beginning,
and that the velocity field is mostly the result of the global dynamics
of the turbulent flow, not the central gravitational action.

As a result of the overall flow, the (2D) ``filament'' that first
appears merges with another structure at $t\sim$~8~Myr (see
Fig.~\ref{nube_evol}).  This results in the jump in properties seen in
Fig.~\ref{nube_evol}, except for $l_J$, which depends on intensive
quantities: the mean temperature and mean density.  Independent of this
merging, at this time $l_J \sim l_x$, suggesting that now the growth of
the cloud is also affected by the gravitational field.

Concerning the time evolution of the energies, it can be seen that the
absolute value of the gravitational energy (long-dashed line) is higher
than the magnetic energy (dotted-dashed  line) by at least two orders
of magnitude.  This indicates that the cloud is supercritical by a
large margin. Furthermore, it can also be seen that the kinetic energy
can not be an agent to stop the collapse: besides being smaller than
$|E_{\rm grav}|$ the velocity field is convergent, implying that it is
not working against gravity.

Finally, we want to stress that the region called the ``molecular
cloud'' (black isocontour at $\rhot = 35$~cm\alamenos 3 in
Figure~\ref{nube_evol}) appears at roughly the same time in which the
region has similar physical and Jeans lengths, i.e., $l_x \sim l_J$,
indicating that the value selected of 35~cm\alamenos 3 is reasonably
indicative of the time at which the cloud becomes self-gravitating.

As a corolary of the fact that the velocity field is not due to the
gravitational potential of the cloud, self-gravitating clouds like that
shown in Figure~\ref{nube_evol} can therefore be produced by the
turbulent velocity field, out of an initially stable medium.

How fast these ``molecular clouds'' can be produced by the general
turbulence must depend on how much mass the streams are carrying, how
strong the compression is, the rate of cooling of the compressed
(shocked) region, the geometry of the compression, etc.
For example, the diffuse structure that it is present when we turned
off the star formation spends $\sim$~7~Myr to become self gravitant.
Nevertheless, Figure~\ref{nube_evol} shows that in the simulations,
self-gravitating structures of 10-20~pc can be coherently formed within
a few Myr by this mechanism. Since the densities and velocities in the
simulations are realistic, this is a plausible mechanism for molecular
cloud formation in the ISM.

\subsection{Comparison with Observations}\label{comparison}

In order to test how the simulations compare with observations, we
construct ``spectra'' for both the densest regions ($\rho \geq \rhot$,
which we call ``molecular cloud'') and for the low-density regions
($\rho < \rhot$,  which we call ``atomic clouds''). Those spectra are
constructed as the mass-weighted velocity (x-component) histograms,
mimicking the emission of an optically thin line observed from the left
hand side of the box, with a spatial resolution of 1.25~pc (1~pixel),
and an ideal telescope. With this approximation, we are implying that
the observed \doceCO\ and H~I line profiles are good representations of
the (mass-weighted) line of sight-velocity field. Also, we are assuming
that the \doceCO\ and H~I emission do not coexist, which is not
necessarily true.

With those spectra (mass-weighted velocity histograms) it is possible
to construct a velocity-position diagram, and compare it with the those
in Fig~\ref{cutposvel}.  Figure~\ref{posvel_code} is the corresponding
velocity-position diagram for the cloud and its surrounding medium
shown in Figure~\ref{nube_evol}i. We emphasize several points of
similarity between this synthetic position-velocity diagram and the
observations (see figure~\ref{cutposvel}).  First, the synthetic CO
emission is always located within a region of strong synthetic H~I
emission; second, there are regions where synthetic H~I emission is
present, but where synthetic \doceCO\ emission is not; third, the
maximum of the high-density emission does not necessarily coincide with
maximum of the low-density emission; fourth, the velocity dispersion is
higher for the synthetic H~I emission than from the synthetic
\doceCO\ emission, and the velocity dispersions of both the low-density
and in the high density gas ``emission'' have similar line-widths to
those of the observational data. Fifth, both \doceCO\ and H~I line
profiles are asymmetric. Finally, as in the case of Taurus, the
filament is almost coherent in velocity dispersion, i.e., it has
approximately the same velocity dispersion along itself, with values of
$\sim$2~\kms\ along 10-20 parsecs.  We will discuss this in \S
\ref{turbcompresions}.

\section{DISCUSSION}\label{discussion}

\subsection{Evidence for Turbulent Compressions}\label{turbcompresions}

It is frequently assumed that the ISM is permeated by small-scale
non-thermal motions that maintain clouds in a stationary state.
Collapse is prevented by those motions, and the role of the external
``intercloud'' medium is only as an agent for (thermal) pressure
confinement and as a shielding for the UV radiation. However, the
external medium also exerts a dynamical influence on the cloud,
compressing and/or disrupting it (Sasao 1973; \cite{hunter79};
\cite{hunt_fleck82}; \cite{hunter_etal86}; \cite{Tohlineetal88};
\cite{VSPP96} \cite{BVS99}). In this sense, the fragmented appearance
of clouds and the existence of multiple velocity components observed at
high velocity resolution suggest that cloud's internal velocity field
possesses a disordered or turbulent component at scales comparable with
the scales of the clouds.  If molecular clouds are embedded in an
intercloud medium of atomic hydrogen (see series of papers by
\cite{Andersson93} and references therein), and if the intercloud
medium is highly turbulent (see e.g.  \cite{Braun99}), it is plausible
that molecular clouds and their surrounding gas are in a dynamical
state.


Based on refinements of the ISM simulations by \cite{PVSP95},
\cite{BVS99} consider the turbulent pressure at cloud
``boundaries''.  They show that this boundary pressure is generally
anisotropic, which distorts the cloud because the energy involved is
generally comparable to the internal cloud kinetic energy.  In their
picture, turbulent motions are not only responsible for cloud support,
as is widely accepted: they may also be responsible through large-scale
modes of the external turbulence, for shaping and compressing the
cloud, possibly even initiating collapse.

In \S \ref{observations}\ we have shown that line-profiles are
asymmetric and show important substructure, features which have been
increasingly noticed in high spectral and spatial resolution
observations (e.g., \cite{Falgaroneetal98}). These features have been
proposed as indicative of large-scale turbulent motions, which may be
shaping, distorting and disrupting the cloud (\cite{BVS99}).

According to the scenario presented here, the observations, which are
also reproduced in the simulations, may be interpreted as follows.
First, because collisions between H~I streams produce the higher
density gas (and for reasons of shielding), there should always be H~I
spatially correlated with molecular gas.  Second, the larger velocity
dispersion of the H~I gas is interpreted as a consequence of its larger
spatial extension, and the compression (shocks) and consequent kinetic
energy dissipation which occurs when the H~I streams collide to form
the cloud.  Finally, because it is the convergence of macro-turbulent
H~I streams that produce the high-density material (see Elmegreen
1996), the H~I line profiles should be asymmetric and frequently
shifted in velocity with respect to the molecular gas\footnote{Note
that in a laminar 2-streams collision producing a shock-bounded slab,
the optically thin line-profile from the `external' medium must be a
double peaked line, at velocities $-v$ and $+v$, where $2 v$ is the
velocity difference between the streams.  On the other hand, the
line-profile from the compressed region must be a single peaked line at
zero velocity. Nevertheless, in a turbulent medium, with density and
velocity fluctuations over the some mean density and velocity profiles
increasing at the center of the compressed region, the double peaked
line-profile might become a single broader line-profile (with its
substructure) enveloping the narrower line-profile that comes from the
compressed region, as its observed from the synthetic line-profiles and
as it is suggested from the observed line-profiles.}.

Thus, the breadth and asymmetry of the H~I line profiles
indicate a macroscopically-turbulent medium in which clouds exchange
mass, momentum and energy with their surroundings. Note that the
picture outlined here does not require a coherent or single triggering
event in which atomic gas pushes molecular gas together as in a SN
explosion.  Our scenario is more general, reflecting statistical
fluctuations resulting from the combined effects of differing sites.

Recognition of these features is not new. The correspondence between
H~I and CO in position-velocity diagrams has already been noted by
several authors (e.g. \cite{BlitzThaddeus80};
\cite{ElmegreenElmegreen87}; see also \cite{Blitz87}; and references
therein). Andersson (1993 and references therein), who observed the
spatial transition between molecular and atomic gas in 62 edges in 14
clouds, found all five observational features that we described at the
end of \S \ref{observations} and \S \ref{comparison}.
Moriarty-Schieven, Andersson \&\ Wannier (1997) suggested that there is
morphological evidence that nearby clouds (L~1457, see cut 6 in
Figs.~\ref{HICO} and \ref{cutposvel}~f) have suffered compressions at
scales similar to the scales of the cloud. These authors proposed an
unobserved SN explosion as a possible mechanism of producing this
morphology. However, the observed morphology is hard to reconcile with
a single point source of compression, while it is fully consistent with
our scenario. Furthermore, the existence of multi-peak, asymmetric
line-profiles at all scales (for larger, see e.g., \cite{Mizunoetal95};
for smaller, see e.g., \cite{Falgaroneetal98}) strongly suggests the
multi-scale nature of the turbulent motions (\cite{BVS99}). Finally,
the velocity-dispersion coherence along large structures have already
seen before both in low-mass clouds (Taurus, \cite{Mizunoetal95}) as
well as in high mass clouds (W51 \cite{CarpenterSanders98}).

\subsection{Limitations of the Model}\label{limitations}

One of the most important limitations of these simulations is that the
cooling laws and the available resolution limit the simulations to
densities $\lesssim $ 100 cm\alamenos 3, so that the evolution of
higher-density (\diezala 2 or more) structures cannot be followed.
This is important, because molecular gas probably cannot be formed
sufficiently rapidly within our density range.  Characteristic $H_2$
formation rate on grains are of the order of $\tau^{-1} \sim n R$,
where $n$ is the density of atomic hydrogen and $R$, the rate
coefficient, has typical values of the order of 3 $\times$ \diezala{-17}
cm\ala 3 seg\alamenos 1 (\cite{Jura75}). Then, the typical $H_2$
formation timescales are given by
\begin{equation} 
\biggl( {\tau_{H_2}\over {\rm yr}} \biggr) \sim 10^6 \biggl( {n\over
10^3 {\rm cm}^{-3}}\biggr)^{-1}.
\label{time_h2}
\end{equation} 
Thus, at our threshold density of 35 cm\alamenos 3, the $H_2$ formation
timescale is 30 Myr, uncomfortably long.

The low values of the density reached in the simulations' clouds
are consequence of the spatial resolution and the mass diffusion,
which smoothes the strong density gradients (see e.g., \cite{VBR97}).
Nevertheless, we envisage that the clouds simulated here would really
collapse to structures of much higher density
in even shorter times (for example, at $n \sim$~100~cm\alamenos 3,
the free-fall time is approximately 3~Myr),
given the strength of the external flows, if we were to use higher
spatial resolution. In this way our ``molecular gas'' above the threshold
density merely indicates the approximate location and velocity of the
higher-density material which would be formed within it. In the real
interstellar gas, there is no barrier to compressing gas in the cloud
to typical Taurus densities of the order of \diezala 3 cm\alamenos 3
so that the molecular gas could be produced as rapidly as the dynamical
compression occurs.

On the other hand, regions that we identify with molecular gas are not
really isothermal in the simulations, but this is not a reason for
concern, since we are not dealing with the internal structure of these
clouds which are at the limit of the resolution anyway.

Another limitation of the simulations is that they are two-dimensional
calculations which assume no variation of properties perpendicular to
the galactic plane.  While colliding streams in 2D produce filaments,
in 3D they may produce sheets.  Nevertheless, the most probable
situation in one in which the colliding streams are oblique, such that
the region that is compressed is a filament again. Also,
inhomogeneities in the physical properties of the streams (density,
temperature, etc.) and a set of  instabilities which are produced when
streams are colliding may destroy the possible appareance of the
sheet.  Furthermore, the structure in 3D simulations is still strongly
filamentary (see e.g., Ostriker et al. 1998; Padoan \&\ Nordlund 1999;
Mac Low 1999; Pichardo et al. 1999). \cite{BVS99} discuss other
possible differences of 2D and 3D simulations.

Thirdly, the energy input into the ISM is assumed to be from high mass
star formation, which seems to be of little relevance to low-mass star
forming regions like Taurus.  However, the only role of the energy
input in the present situation is to produce large-scale flows and
turbulence which in turn produce the ``molecular cloud''; since the
predicted motions agree quite well with the H~I data, our approach is
justified.  Moreover, since the stellar energy input is turned off to
follow cloud formation, disruption by local high-mass stars is not an
issue.

It should be emphasized that while we are considering the simulations
with a view toward understanding the Taurus region, we are not
attempting to model the complex in detail, because turbulent flows
exhibit a chaotic behavior, i.e., the time histories of arbitrarily
close initial conditions diverge exponentially and end up completely
different (e.g. \cite{Lessieur90}). Instead, they are only expected to
statistically reproduce the dynamical relationship between H~I flows
and molecular gas in a statistical sense

\section{SUMMARY}\label{conclusions}

In this paper we have used H~I and \doceCO\ data to show that the
dynamical features observed through Taurus can be interpreted as
large-scale compressions of the atomic gas producing the molecular gas
in timescales of few Myr. Also, we followed the evolution of an small
piece of 2D numerical MHD simulations to show how large scale
turbulence is able to trigger collapse in  density structures that
originally are Jeans stable. The timescales for doing this also are few
Myr, suggesting that low-density H~I streams can be compressing and
producing the high density regions in Taurus.

With this picture in mind, we suggest that the Taurus molecular cloud
may have formed quite recently ($\sim$~3~Myr ago). This would solve the
post T-Tauri problem, suggesting that the very few stars with ages
between 5 and 10~Myr in the region might be field stars.
%
%
We also suggest that the general large-scale interstellar turbulence
(in H~I) is the mechanism responsible for triggering coherent collapse
in regions which apparently are dynamically disconnected.

In this paper we have only considered the timescale of cloud formation,
not the timescale of disruption or dissipation.  Even with a rapid
formation time, if clouds live for 10 Myr or more, at least some
complexes should have PTTSs.  Thus, our investigation only addresses
one-half of the more general post-T Tauri problem.  We intend to
explore the rate of cloud disruption and dissipation in a future
contribution.

\acknowledgements

The simulations were performed on the Cray Y-MP/4-64 of DGSCA, UNAM.
This work has received partial support from grants UNAM/DGAPA IN105295
and UNAM/CRAY SC-008397 to E.V.-S.\ and a UNAM/DGAPA fellowship to
J.B.-P. Cesar \&\ John: discusiones. Ungerechts, datos.

{}

\begin{figure}
\plotone{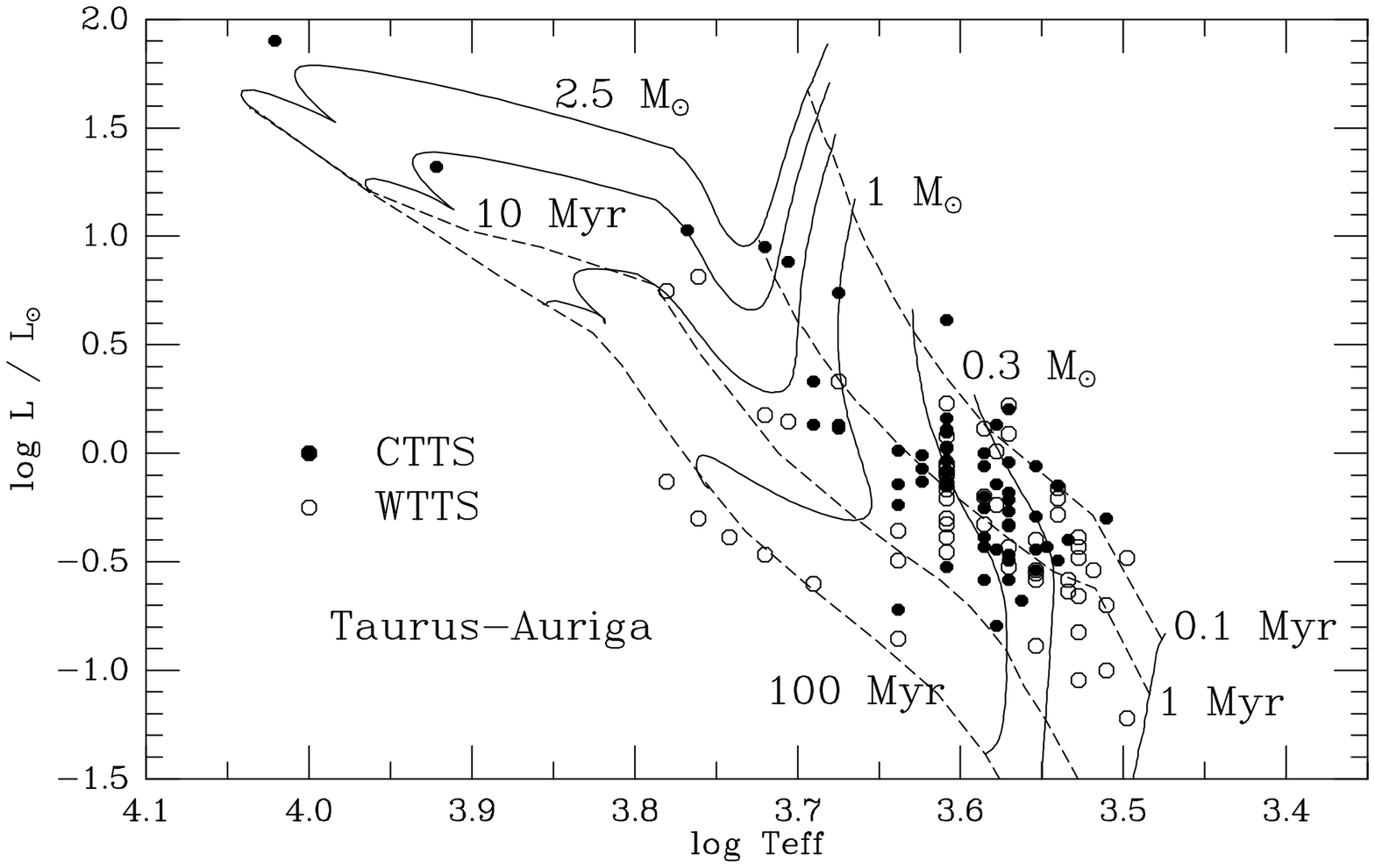}
\end{figure} 

\begin{figure}
\plotone{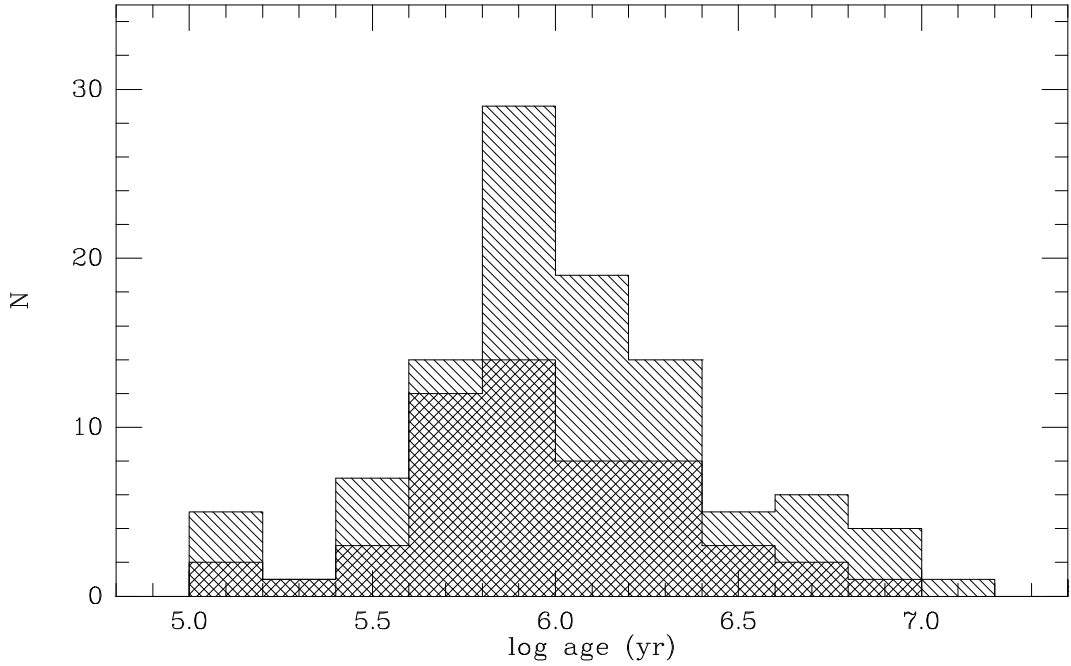}
\caption{ a) HR Diagram for stars in the Taurus-Auriga molecular cloud
region, with evolutionary tracks from D'Antona \&\  Mazziteli (1994).
Filled circles denote the Classical T Tauri Stars (CTTS). Open circles
denote the Weak T Tauri Stars (WTTS).  The great majority of stars have
ages of $\sim$~1~Myr, and only a few stars have ages greater than
$\sim$~3~Myr. The group of stars near the $\sim$~100~Myr isocrone were
discovered with Einstein X-ray observations (Walter et al. 1988); X-ray
detected sources in the region are generally older than 10 Myr
(\cite{Bricenoetal97}), and are likely to be field stars.  b)  Age
histogram for the stars in fig~1a. Single line denotes the classical
T-Tauri stars. Crossed lines, the weak T-Tauri stars. Note that the
great majority of stars is between 1 and 3~Myr.
\label{figuralee}}
\end{figure} 
\newpage

\begin{figure}
\plotone{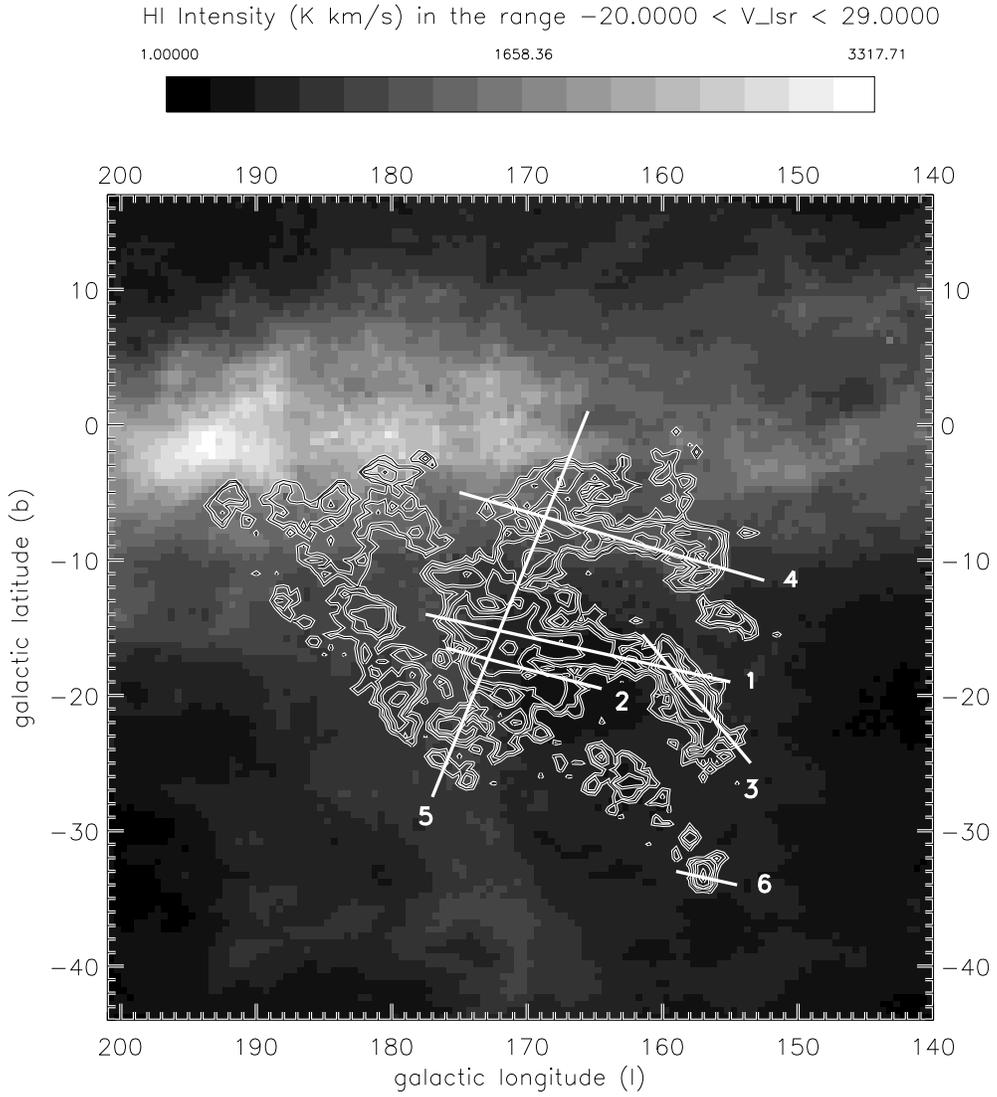}
\caption{Velocity integrated (from $-20.3$ to 29.08~\kms) large-scale map
of antenna temperature for CO (\cite{UT87}) and HI (\cite{dap97}) data
through Taurus, in units of K \kms. Contours:  2, 4, 8, 16, 32, 64. The
grey scale is indicated on the top. The white lines denote the places
where the velocity-position diagrams have been made. (Lee: I have to
include also the cut in L1457)
\label{HICO}}
\end{figure}

\begin{figure}
\plotone{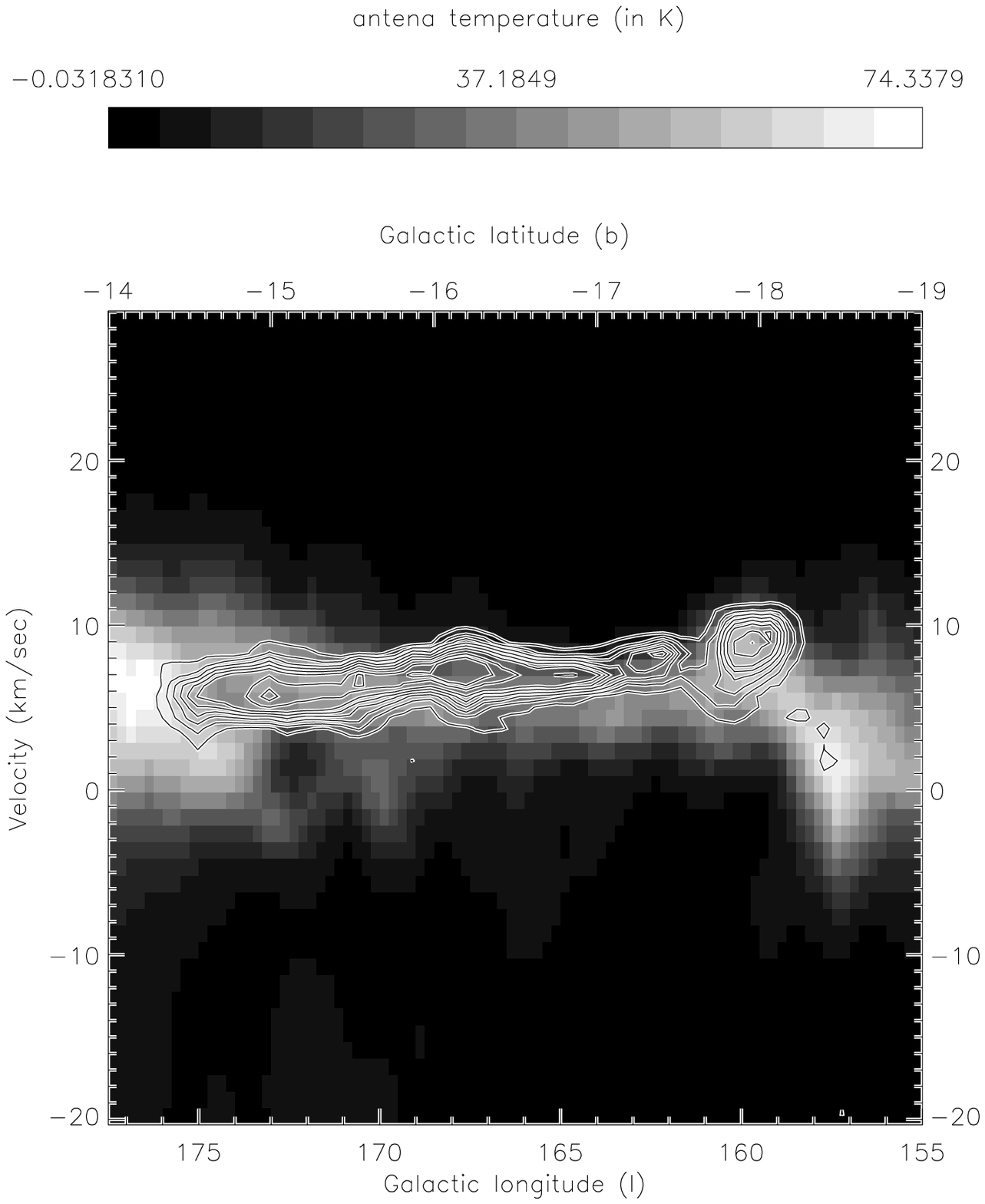}
\end{figure}

\begin{figure}
\plotone{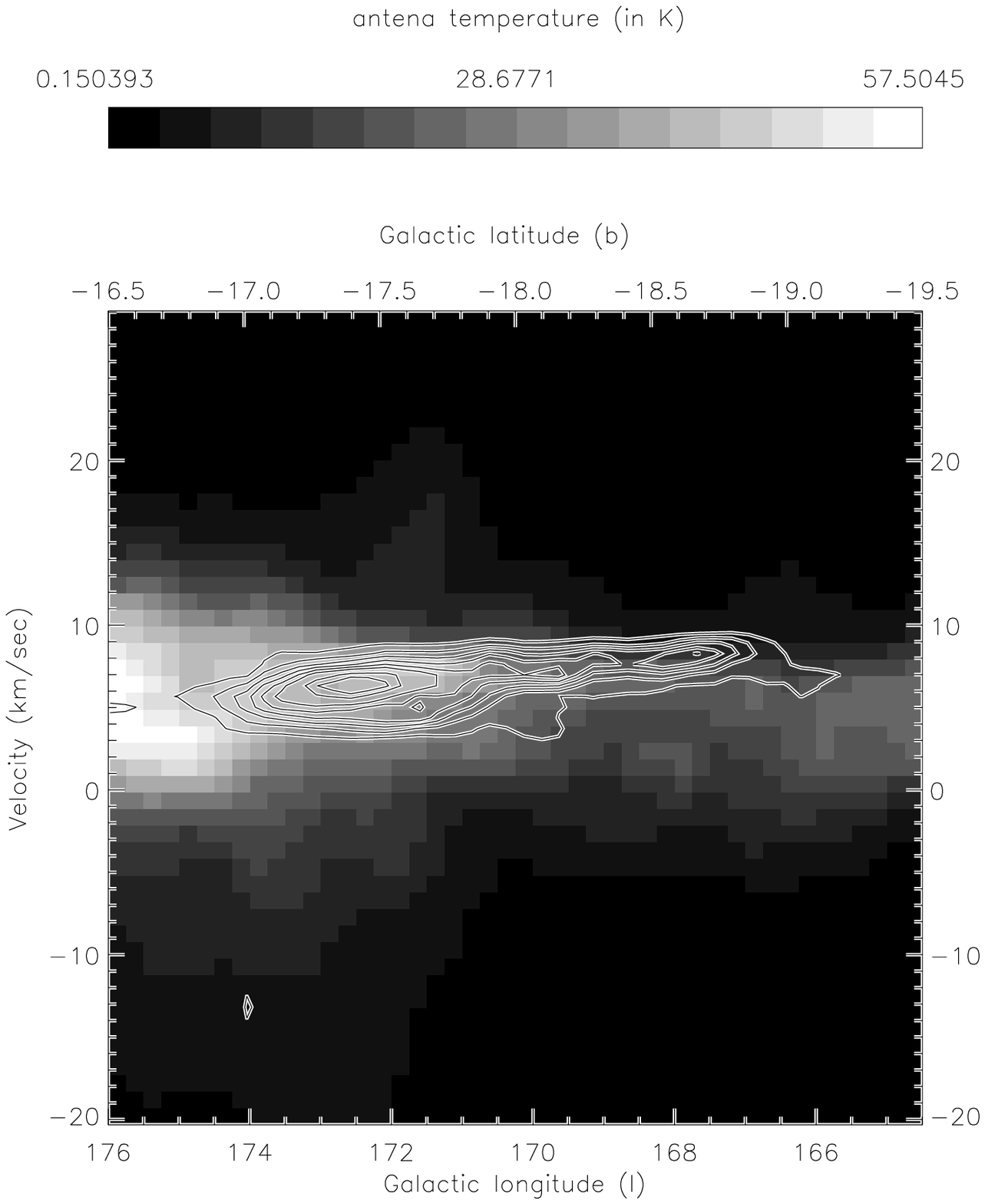}
\end{figure}
 
\begin{figure}
\plotone{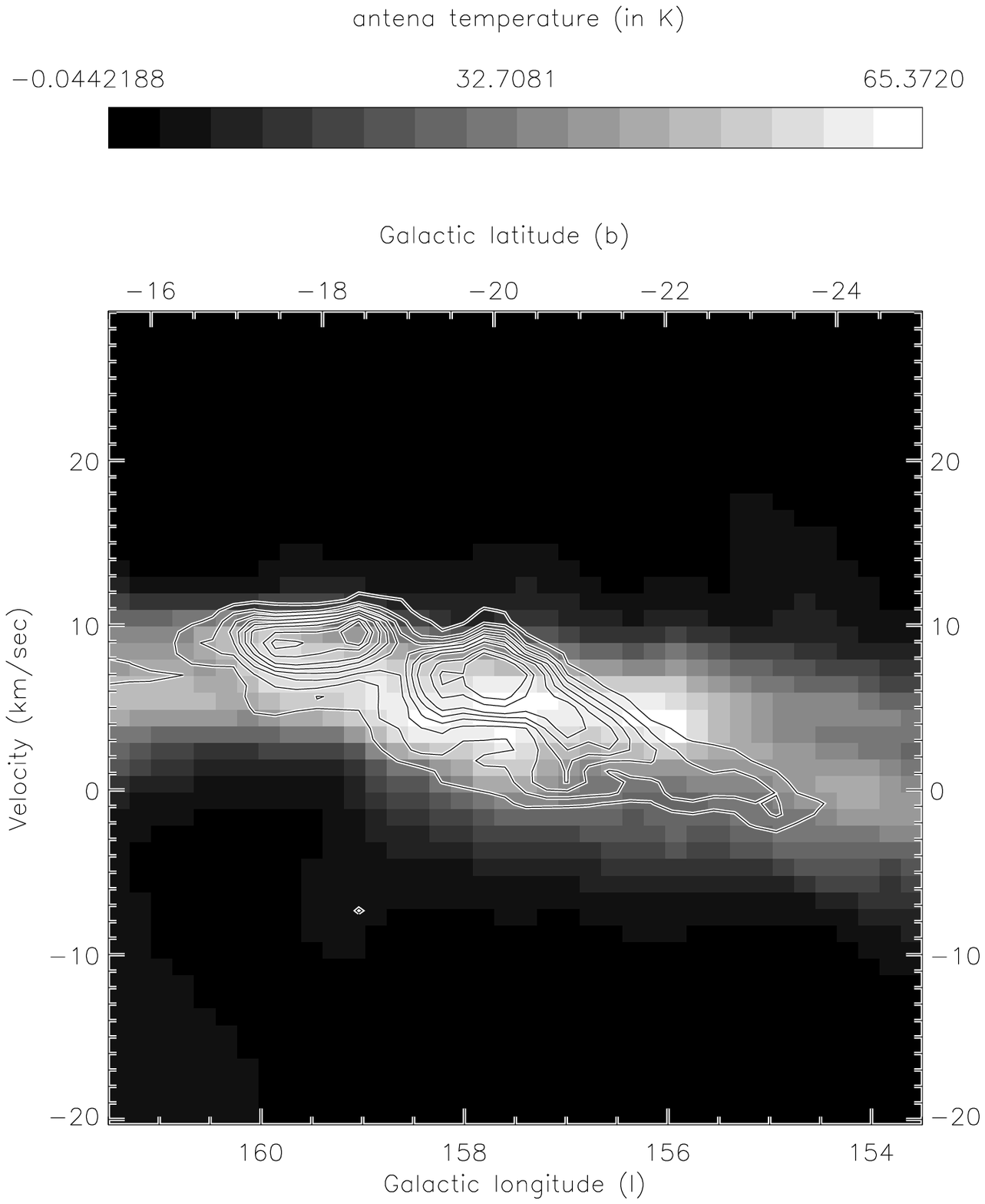}
\end{figure}

\begin{figure}
\plotone{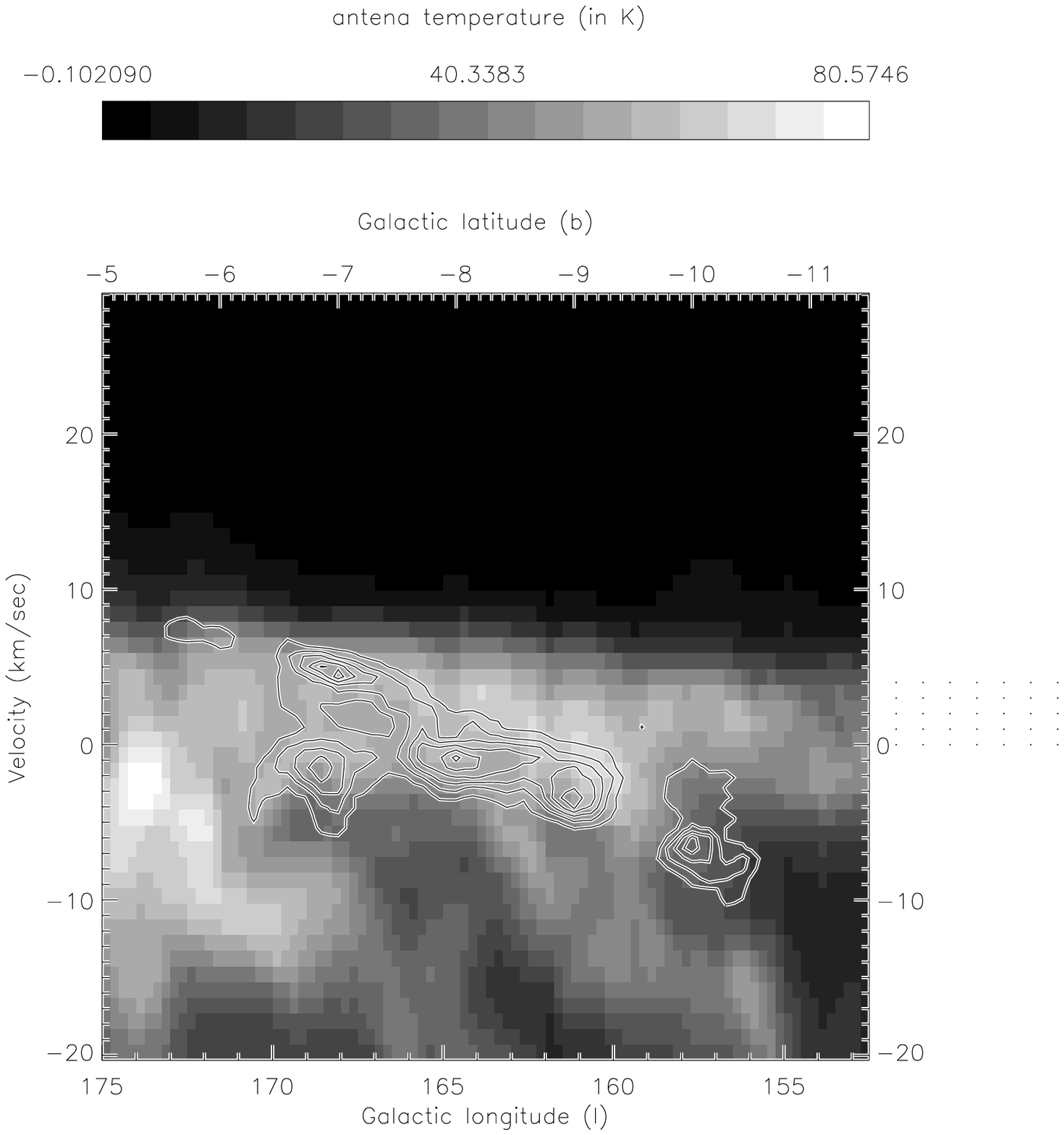}
\end{figure}
 
\begin{figure}
\plotone{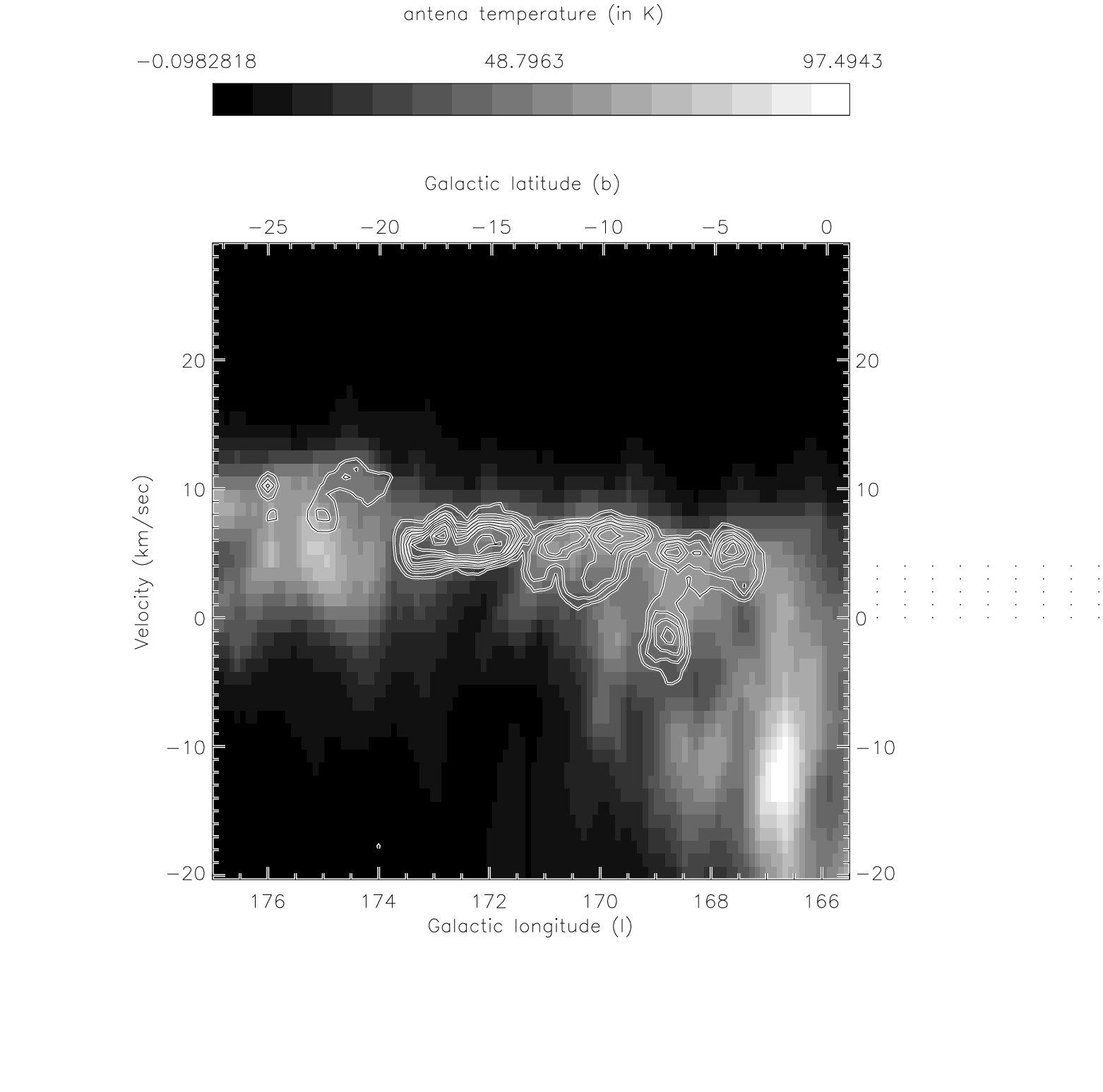}
\end{figure}

\begin{figure}
\plotone{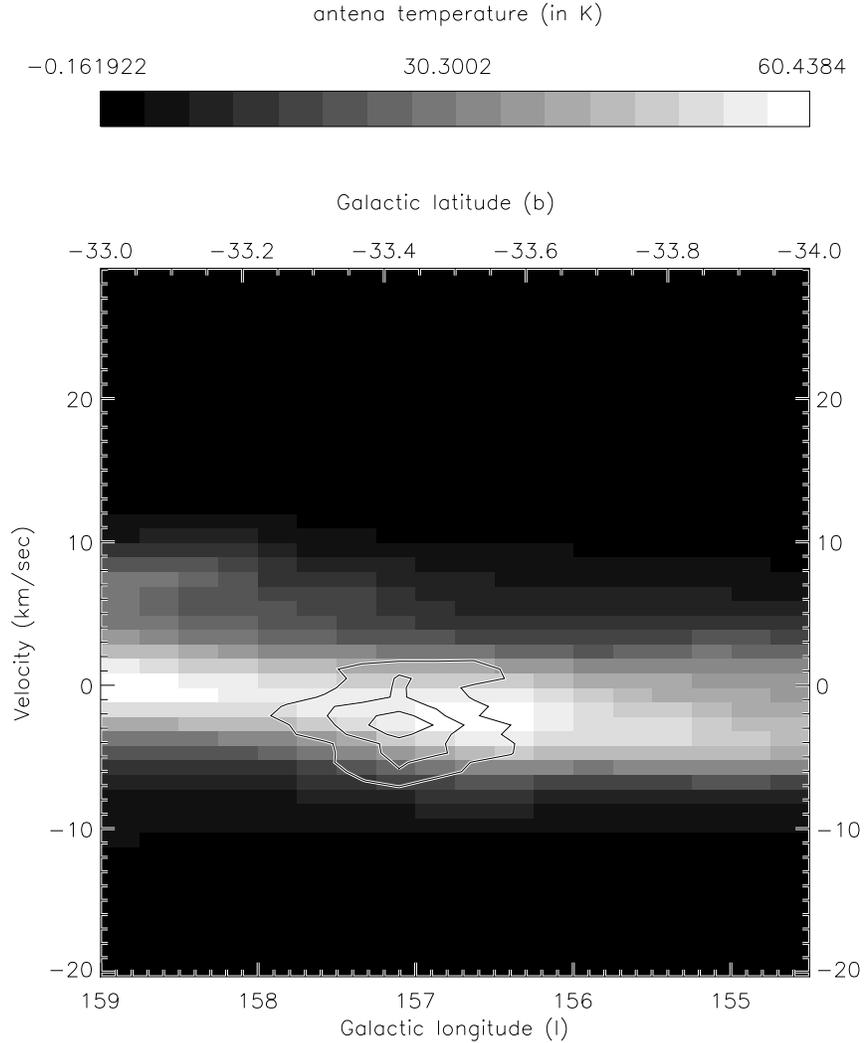}
\caption{Velocity-Position diagrams at the positions showed in
fig.~\ref{HICO}. Grey scale denotes the HI emission, and contours denote
the \doceCO\ emission. The $y$-axis denotes the velocity in \kms; the
upper $x$-axis denotes the galactic latitude, and the lower
$x$-axis denotes the galactic longitude, both in degrees. Note 
the following general
characteristics:  a) wherever there is a \doceCO, HI is also found,
with an approximate column density similar to that required by
shielding.  b) the converse is not true: not all HI in this velocity
system is associated with molecular gas.  c) at the same spatial
position, the HI emission often does not peak at the same velocity than
the \doceCO; frequently there is a shift of a few \kms between the two
species. d) velocity widths in the HI spectra are larger than the
\doceCO\ spectra by a factor of roughly 3 or more.  e) both the
\doceCO\ and HI line profiles are asymmetric, as indicated by the
variation of gray-scale and contours.
\label{cutposvel}}
\end{figure}

\begin{figure}
\plotone{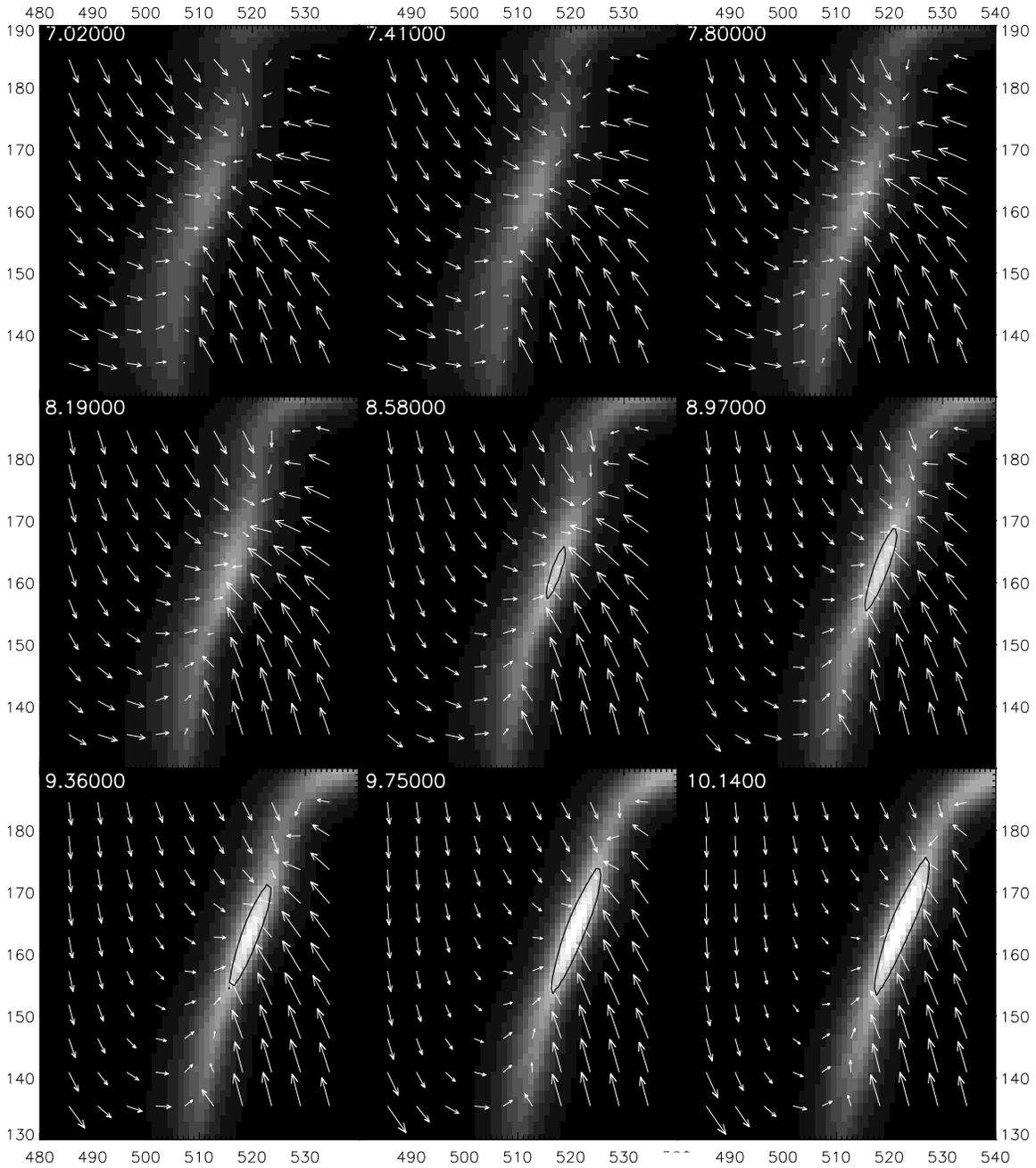}
\vskip 1cm
\caption{Time sequence of a small region of the simulation. We show the
density field (grey scale), ranging from $\rho < 0.31$~cm\alamenos 3
(darker) to $\rho >$~56~cm\alamenos 3. The label in the upper left
corner in each panel is the age in Myr after turning-off the star
formation in the simulation. Note that material with $n >
35$~cm\alamenos 3 (the ``molecular cloud'') is formed in few Myr by the
large-scale compression.
\label{nube_evol}}
\end{figure}

\begin{figure}
\plotone{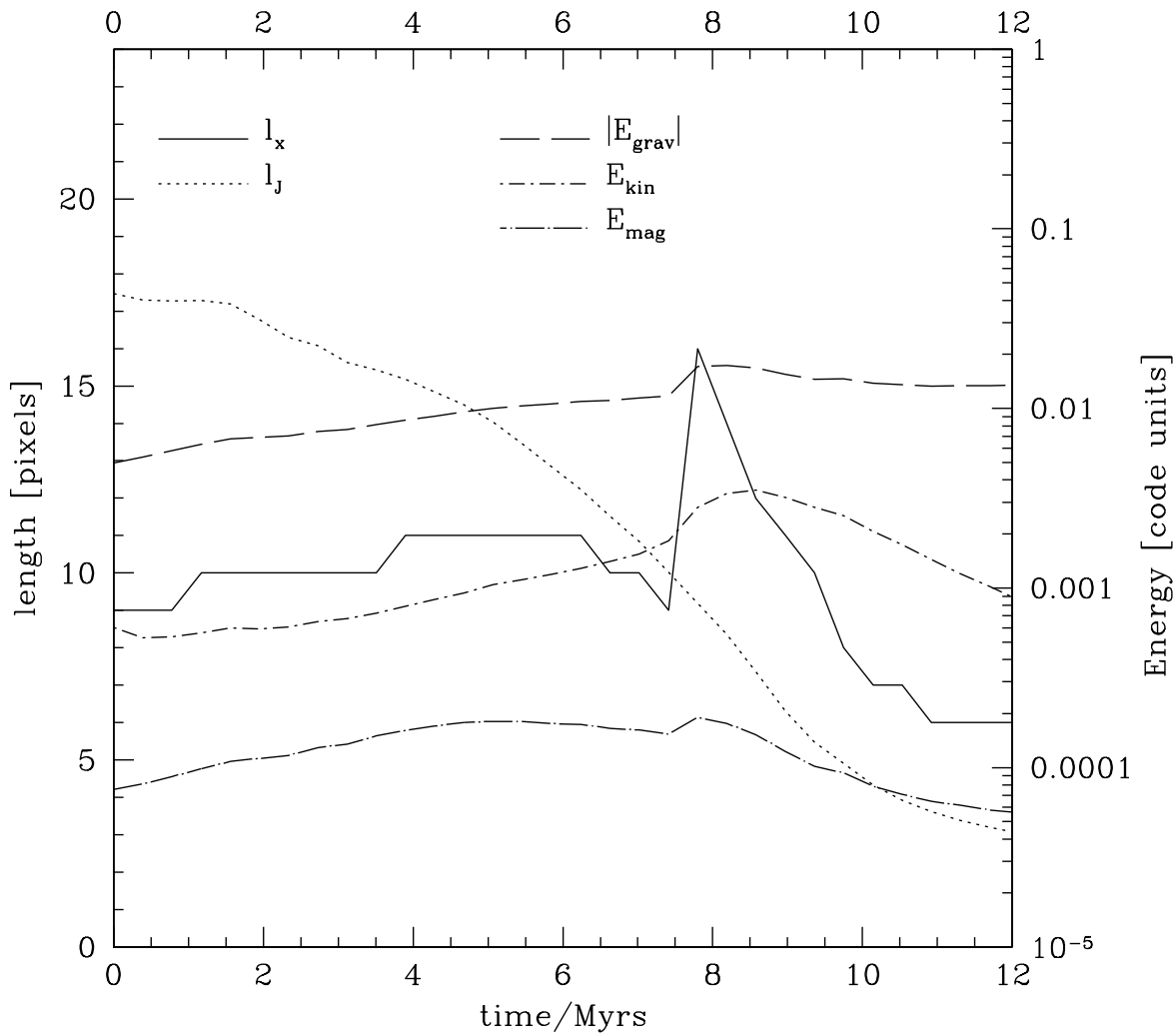}
\caption{Time-evolution of the characteristic lengths (left-hand side
of the $y$-axis), and energies ($x$-axis) for the region defined as a
connected set of pixels with density equal or higher than 1/2 value of
the density maxima at each timestep. Solid line, $l_x$, the maximum
length in the $x$-direction; dotted-line, $l_J$, the Jeans length as in
eq. (\ref{jeans}). Short-dashed line, absolute value of the
gravitational energy; long-dashed line, kinetic energy in the frame of
the cloud, and dotted-dashed line, magnetic energy. Note that initially
the cloud is Jeans-stable, but after some time, it becomes
Jeans-unstable. Also, note that the magnetic energy is two orders of
magnitude lower than the gravitational energy, suggesting that there is
no magnetic-flux problem, i.e., the cloud is supercriticall. The strong
jump in the extensive quantities at $t\sim 8$~Myr. is due to the
merging of the filament with another density feature, as can be seen in
Fig.~\ref{nube_evol}.
\label{longitudes}}
\end{figure}

\begin{figure}
\plotone{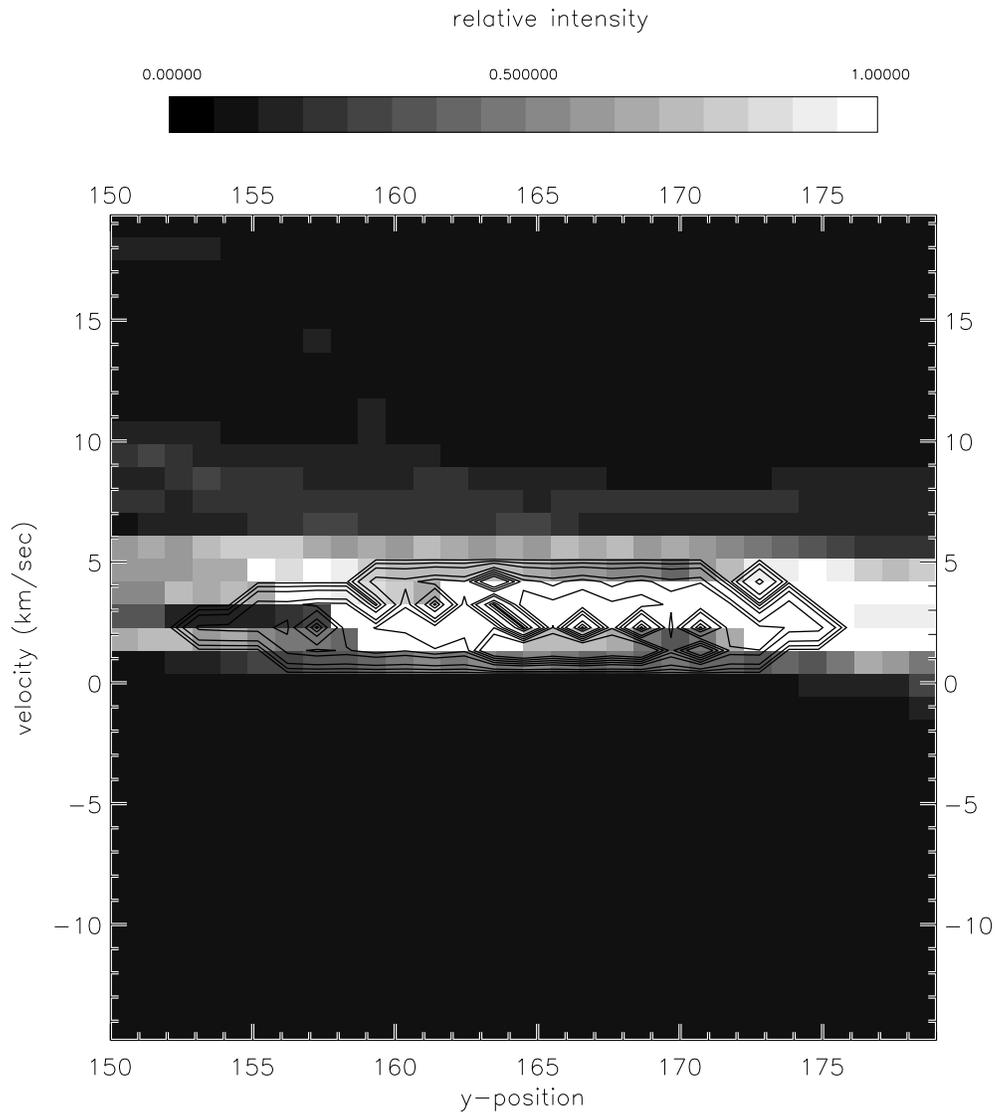}
\caption{Synthetic Velocity-Position diagram for the cloud in panel
\ref{nube_evol}i. As in fig.~\ref{cutposvel}, grey scale denotes the ``HI
emission'', and contours denotes the ``emission'' coming from the
``molecular cloud''.  Note that the same characteristics observed in
the observational position-velocity diagrams are reproduced also here.
\label{posvel_code}}
\end{figure}

\end{document}